\documentclass[12pt]{article}
\usepackage{amssymb}
\usepackage{amsmath}
\def\d{\delta}

\def\be{\begin{equation}}
\def\ee{\end{equation}}
\def\arr{\begin{array}{rll}}
\def\ea{\end{array}}
\def\bea{\begin{eqnarray}}
\def\eea{\end{eqnarray}}
\def\N2{$N{=}2$}

\def\>{\rangle}
\def\<{\langle}
\def\+{\dagger}
\def\={\ =\ }

\begin{document}
\renewcommand{\thefootnote}{\arabic{footnote}}
\begin{titlepage}
\setcounter{page}{0}
\vskip 1cm
\begin{center}
$\qquad$\\
\vskip 3cm
{\LARGE\bf $\mathcal{N}=2$ supersymmetric}\\
\vskip 0.5cm
{\LARGE\bf Pais-Uhlenbeck oscillator}\\
\vskip 1cm
$
\textrm{\Large Ivan Masterov\ }
$
\vskip 0.7cm
{\it
Laboratory of Mathematical Physics, Tomsk Polytechnic University, \\
634050 Tomsk, Lenin Ave. 30, Russian Federation}
\vskip 0.7cm
{E-mail: masterov@tpu.ru}

\end{center}
\vskip 1cm
\begin{abstract} \noindent
We construct an $\,\mathcal{N}=2$ supersymmetric extension of the Pais-Uhlenbeck oscillator for distinct frequencies of oscillation. A link to a set of decoupled $\,\mathcal{N}=2$ supersymmetric harmonic oscillators with alternating sign in the Hamiltonian is introduced. Symmetries of the model are discussed in detail. The investigation of a quantum counterpart of the constructed model shows that the corresponding Fock space contains negative norm states and the energy spectrum of the system is unbounded from below.

\end{abstract}


\vskip 0.5cm


\end{titlepage}

\noindent
{\bf 1. Introduction}
\vskip 0.5cm

The range of physical problems described in terms of higher-derivative systems is wide. These systems appear in such important areas of physics as quantum gravity \cite{Stelle}, string theory \cite{Eliezer}, noncommutative quantum mechanics \cite{Lukierski}-\cite{Kosinski_11}, the theory of particles with rigidity \cite{Plyushchay}. The interest in higher-derivative theories is mostly due to their attractive renormalization properties \cite{Stelle,Thiring} (see also Refs. \cite{Anupam_1,Anupam_2} and references therein). On the other hand, they generically reveal a linear instability in classical dynamics \cite{Ostrogradski} as well as violation of unitarity and/or presence of ghost states when quantum counterparts are considered \cite{Pais,Woodard}. The desire to solve these long standing problems motivates an investigation of the so-called Pais-Uhlenbeck (PU) oscillator \cite{Pais}.

The dynamics of the $2n$-order PU oscillator in arbitrary dimension is governed by the equation of motion
\bea\label{EOM}
\prod_{m=0}^{n-1}\left(\frac{d^2}{dt^2}+\omega_m^{2}\right)x_i(t)=0,
\eea
where $\omega_m$ with $m=0,1,..,n-1$ are frequencies of oscillation. In the original work \cite{Pais}, it has been shown that the PU oscillator is dynamically equivalent to a direct sum of harmonic oscillators with alternating sign in the Hamiltonian. In the oscillator representation it is easy to see that the energy of the PU oscillator can take both positive and negative values. However, the classical dynamics of the PU oscillator is stable. Instabilities can show up when an interaction between the negative and positive energy modes is switched on.

The quantum PU oscillator faces the ghost problem which manifests itself in two different ways \cite{Woodard}. The first one is characterized by the presence of negative norm eigenstates of the Hamiltonian. In this case the energy spectrum is bounded from below, however. In the second description, all eigenstates of the Hamiltonian have positive definite norm but the energy spectrum is unbounded from below and, hence, there is no a ground state.

Recently the PU oscillator has been discussed with regard to dynamical realizations of the so-called $l$-conformal Newton-Hooke algebra \cite{Negro_1}-\cite{Galajinsky_3}. In particular, it has been shown that the PU oscillator enjoys the $l$-conformal Newton-Hooke symmetry provided frequencies of oscillation form the arithmetic sequence $\omega_{k}=(2k+1)\omega_{0}$ \cite{Galajinsky_1,PU} (see also Refs. \cite{Gomis}-\cite{Andrzejewski_1}).

The presence of conformal invariance allows one to construct various supersymmetric generalizations of the PU oscillator by applying appropriate coordinate transformations \cite{Masterov_2}-\cite{Masterov_1}. One of the main motivations to building such extensions was to solve the ghost problem with the aid of the supersymmetry \cite{Smilga}. Indeed, suppose that a quantum system described by a Hermitian Hamiltonian $\hat{H}$ possesses the
supersymmetry charges $\hat{Q}$, $\hat{\bar{Q}}$ which are hermitian conjugates of each other. Then the energy spectrum is nonnegative and a ground state exists provided the (anti)commutation relations
\bea\label{str}
\{\hat{Q},\hat{Q}\}=0,\quad [\hat{H},\hat{Q}]=0,\quad \{\hat{Q},\hat{\bar{Q}}\}=2\hat{H},\quad[\hat{H},\hat{\bar{Q}}]=0,\quad \{\hat{\bar{Q}},\hat{\bar{Q}}\}=0
\eea
hold and a state space of the system contains only positive norm states (see a related discussion in Ref. \cite{Smilga}). An attempt to solve the ghost problem in this way has been made in Ref. \cite{Smilga}, but the supercharges in \cite{Smilga} are not hermitian conjugates of each other.

The purpose of this work is to generalize an $\,\mathcal{N}=2$ supersymmetric extension of the PU oscillator obtained in \cite{Masterov_1} to the case of arbitrary frequencies of oscillation and to study whether or not supersymmetry may help to resolve the ghost problem for this system.

The paper is organized as follows. In Section 2, we introduce an action functional of an $\,\mathcal{N}=2$ supersymmetric PU oscillator. Oscillator coordinates are found in terms of which the model decomposes into the sum of decoupled $\,\mathcal{N}=2$ supersymmetric harmonic oscillators with alternating sign. In Section 3, a quantum version of the $\,\mathcal{N}=2$ supersymmetric PU oscillator is considered. We summarize our results and discuss further possible developments in the concluding Section 4.

\vskip 0.5cm
\noindent
{\bf 2. Classical action}
\vskip 0.5cm

Let us start with the action functional of an $\,\mathcal{N}=2$ supersymmetric extension of the PU oscillator introduced in \cite{Masterov_1}\footnote{Summation over repeated spatial indices $i=1,2,..,d$ is understood, unless otherwise is explicitly stated.}
\bea\label{action1}
\begin{aligned}
&
S=\frac{1}{2}\int d t\left(x_i\prod_{k=0}^{n-1}\left(\frac{d^2}{d t^2}+\frac{(2k+1)^2}{R^{2}}\right)x_i-\,z_i\prod_{k=0}^{n-2}\left(\frac{d^2}{dt^2}+\frac{(2k+1)^2}{R^2}\right)z_i-\right.
\\[7pt]
&
\qquad\qquad\quad-i\psi_i\left(\frac{d}{dt}+\frac{(2n-1)i}{R}\right)\prod_{k=0}^{n-2}\left(\frac{d^2}{dt^2}+\frac{(2k+1)^2}{R^2}\right)
\bar{\psi}_i-
\\[7pt]
&
\qquad\qquad\quad\left.-i\bar{\psi}_i\left(\frac{d}{dt}-\frac{(2n-1)i}{R}\right)\prod_{k=0}^{n-2}\left(\frac{d^2}{dt^2}+\frac{(2k+1)^2}{R^2}\right)
\psi_i\right),
\end{aligned}
\eea
where $x_i$ are bosonic coordinates, $\psi_i$ and $\bar{\psi}_i$ are fermionic coordinates which are
complex conjugates of each other, $z_i$ are extra bosonic coordinates, and $R$ is an arbitrary real constant\footnote{In \cite{Masterov_1} $\frac{1}{R^2}$ played the role of a cosmological constant \cite{Gibbons} (see also \cite{Galajinsky_6,Galajinsky_4}).}. This action is invariant under transformations which form $\,\mathcal{N}=2$, $l=\frac{2n-1}{2}$-conformal Newton-Hooke supergroup \cite{Masterov_1,Masterov_3}. Let us focus on the supersymmetry transformations of the action (\ref{action1}) which have the form \cite{Masterov_1}
\bea\label{transform}
\begin{aligned}
&
\d x_i=\psi_i\alpha+\bar{\psi}_i\bar{\alpha},\quad\;\d z_i=\left(i\dot{\psi}_i+\frac{2n-1}{R}\psi_i\right)\alpha+\left(-i\dot{\bar{\psi}}_i+\frac{2n-1}{R}\bar{\psi}_i\right)\bar{\alpha},
\\[3pt]
&
\d\psi_i=\left(-i\dot{x}_i+\frac{2n-1}{R} x_i-z_i\right)\bar{\alpha},\qquad \d\bar{\psi}_i=\left(-i\dot{x}_i-\frac{2n-1}{R} x_i+z_i\right)\alpha,
\end{aligned}
\eea
where $\alpha$ and $\bar{\alpha}$ are odd infinitesimal parameters and $\dot{x}_i\equiv\frac{d x_i}{dt}$.

As was mentioned above, the restrictions on frequencies of oscillation in the model (\ref{action1}) come from the requirement of conformal invariance. In this paper we choose to abandon the conformal invariance and generalize the model (\ref{action1}) to the case of arbitrary frequencies
\bea\label{act}
\begin{aligned}
&
S=\frac{1}{2}\int\,dt\left(x_i\prod_{k=0}^{n-1}\left(\frac{d^2}{dt^2}+\omega_k^{2}\right)x_i-
i\psi_i\prod_{k=1-n}^{n-1}\left(\frac{d}{dt}+i\omega_k\right)\bar{\psi}_i-\right.
\\[3pt]
&
\left.\qquad\qquad\;\,-i\bar{\psi}_i\prod_{k=1-n}^{n-1}\left(\frac{d}{dt}-i\omega_k\right)\psi_i-
z_i\prod_{k=1}^{n-1}\left(\frac{d^2}{dt^2}+\omega_k^{2}\right)z_i\right),
\end{aligned}
\eea
where we denote $\omega_{-k}=-\omega_k$. An analogue of the transformations (\ref{transform}) for the action (\ref{act}) reads
\bea\label{tr}
\left.
\begin{aligned}
&
\d x_i=\psi_i\alpha+\bar{\psi}_i\bar{\alpha},\qquad \d z_i=\left(i\dot{\psi}_i+\omega_0\psi_i\right)\alpha+ \left(-i\dot{\bar{\psi}}_i+\omega_0\bar{\psi}_i\right)\bar{\alpha},\\[5pt]
&
\;\;\;\d\psi_i=\left(-i\dot{x}_i+\omega_0 x_i-z_i\right)\bar{\alpha},\qquad\d\bar{\psi}_i=\left(-i\dot{x}_i-\omega_0 x_i+z_i\right)\alpha.
\end{aligned}
\right.
\eea

The so-called oscillator coordinates, which proved to be very useful in describing the bosonic limit of the model (\ref{act}), have been introduced in \cite{Pais}. They involve the differential operator which enters the dynamical equation itself with a slight modification. Given the equation of motion (\ref{EOM}), one defines the oscillator coordinates to be
\bea\label{coord1}
x_i^{k}=\prod_{m=0\atop m\neq k}^{n-1}\left(\frac{d^2}{dt^2}+\omega_{m}^{2}\right)x_i,\qquad k=0,1,..,n-1.
\eea
Taking into account other equations of motion
\bea
\prod_{m=1-n}^{n-1}\left(\frac{d}{dt}-i\omega_m\right)\psi_i=0,\qquad \prod_{m=1}^{n-1}\left(\frac{d^2}{dt^2}+\omega_m^{2}\right)z_i=0, \qquad \prod_{m=1-n}^{n-1}\left(\frac{d}{dt}+i\omega_m\right)\bar{\psi}_i=0,
\eea
one can define similar oscillator coordinates for the remaining variables
\bea\label{coord2}
\begin{aligned}
&
\psi_i^{k}=\prod_{m=1-n\atop m\neq k}^{n-1}\left(\frac{d}{dt}-i\omega_m\right)\psi_i,\qquad k=-n+1,-n+2,..,n-1;
\\[3pt]
&
x_i^{-k}=\prod_{m=1\atop m\neq k}^{n-1}\left(\frac{d^2}{dt^2}+\omega_{m}^{2}\right)z_i,\qquad k=1,2,..,n-1;
\\[3pt]
&
\bar{\psi}_i^{k}=\prod_{m=1-n\atop m\neq k}^{n-1}\left(\frac{d}{dt}+i\omega_m\right)\bar{\psi}_i,\qquad k=-n+1,-n+2,..,n-1.
\end{aligned}
\eea
It should be noted that, since the coordinates $\psi_i$ and $\bar{\psi}_i$ are complex conjugates of each other, so are the new variables $\psi_i^{k}$ and $\bar{\psi}_i^{k}$ with $k=-n+1,-n+2,..,n-1$.
The new variables (\ref{coord1}) and (\ref{coord2}) obey to the following equations of motion:
\bea\label{EOM1}
\ddot{x}_i^{k}+\omega_{k}^{2}x_i^{k}=0,\qquad \dot{\psi}_i^{k}-i\omega_{k}\psi_i^{k}=0,\qquad \dot{\bar{\psi}}_i^{k}+i\omega_k\bar{\psi}_i^{k}=0,\;\,
\eea
where $k=-n+1,-n+2,..,n-1$.

At the next step, let us rewrite the action functional (\ref{act}) in terms of the new variables (\ref{coord1}) and (\ref{coord2}). It is natural to search the action in the form
\bea\label{act1}
S=\frac{1}{2}\int dt \sum_{k=-n+1}^{n-1}\left(\rho_{k}x_i^{k}(\ddot{x}_i^{k}+\omega_{k}^{2}x_i^{k})+
i\beta_{k}\psi_i^{k}(\dot{\bar{\psi}}_i^{k}+i\omega_k\bar{\psi}_i^{k})+i\bar{\beta}_k\bar{\psi}_i^{k}(\dot{\psi}_i^{k}-i\omega_k\psi_i^{k})\right),
\eea
where $\rho_k$, $\beta_{k}$, and $\bar{\beta}_k$ are constants to be found. Comparing (\ref{act}) and (\ref{act1}), one derives the restrictions on the coefficients
\bea
\sum_{k=0}^{n-1}\rho_k x_i^{k}=x_i,\qquad \sum_{k=-n+1}^{-1}\rho_{k}x_i^{k}=-z_i,\quad \sum_{k=-n+1}^{n-1}\beta_k\psi_i^{k}=-\psi_i,\quad \sum_{k=-n+1}^{n-1}\bar{\beta}_k\bar{\psi}_i^{k}=-\bar{\psi}_i.
\eea
Each of these expressions turns out to be equivalent to a system of linear algebraic equations which has a unique solution provided all frequencies are different. Apart from $\omega_0$ which is allowed to take zero value, all other frequencies are assumed to be nonzero (for the details see Appendix A). A straightforward calculation yields
\bea\label{coef}
\begin{aligned}
&
\rho_k=\left\{
\begin{aligned}
&
\frac{(-1)^{k}}{\prod\limits_{i_1=0}^{k-1}(\omega_{k}^{2}-\omega_{i_1}^{2})\prod\limits_{i_2=k+1}^{n-1}(\omega_{i_2}^{2}-\omega_{k}^{2})},\qquad\quad\;\;\, k=0,1,..,n-1,
\\[10pt]
&
\frac{(-1)^{k}}{\prod\limits_{i_1=1}^{-k-1}(\omega_{k}^{2}-\omega_{i_1}^{2})\prod\limits_{i_2=-k+1}^{n-1}(\omega_{i_2}^{2}-\omega_{k}^{2})},\qquad\;\;\, k=-n+1,-n+2,..,-1;
\end{aligned}
\right.
\\[8pt]
&
\beta_{k}=\bar{\beta}_k=\frac{(-1)^{k+1}}{\prod\limits_{i_1=-n+1}^{k-1}(\omega_{k}-\omega_{i_1})\prod\limits_{i_2=k+1}^{n-1}(\omega_{i_2}-\omega_{k})},\quad k=-n+1,-n+2,..,n-1.
\end{aligned}
\eea
If, for definiteness, we choose $0\leq\omega_{0}<\omega_{1}<...<\omega_{n-1}$, then the denominators of all the fractions on the right hand side of Eqs. (\ref{coef}) are positive. Subsequent redefinition of the variables
\bea\label{redef}
x_i^{k}\,\rightarrow\,\sqrt{|\rho_{k}|}x_i^{k},\qquad \psi_i^{k}\,\rightarrow\,\sqrt{|\beta_k|}\psi_i^{k}, \qquad \bar{\psi}_i^{k}\,\rightarrow\,\sqrt{|\bar{\beta}_k|}\bar{\psi}_i^{k},
\eea
and discarding total derivative terms yields the following action functional:
\bea\label{act2}
S=\frac{1}{2}\int dt\sum_{k=-n+1}^{n-1}(-1)^{k+1}\left(\dot{x}_i^{k}\dot{x}_i^{k}-\omega_k^{2}x_i^{k}x_i^{k}+i\psi_i^{k}\dot{\bar{\psi}}_i^{k} +i\bar{\psi}_i^{k}\dot{\psi}_i^{k}-2\omega_k\psi_i^{k}\bar{\psi}_i^{k}\right).
\eea
The action describes a set of decoupled $\,\mathcal{N}=2$ supersymmetric harmonic oscillators which alternate in sign. This correlates with the analysis in \cite{Pais} for the corresponding bosonic part of the full model.

\vskip 0.5cm
\noindent
{\bf 3. Quantization}
\vskip 0.5cm

Let us consider the Hamiltonian formulation for the model (\ref{act2}). It relies upon the graded Poisson bracket\footnote{When analyzing (\ref{act2}) within the Hamiltonian formalism, one reveals fermionic second class constraints. Resolving these constraints, one can remove momenta canonically conjugate to $\psi_i^{k}$ and $\bar{\psi}_i^{k}$. The corresponding Dirac bracket yields (\ref{Poisson}).}
\bea\label{Poisson}
[A,B\}=\sum_{k=-n+1}^{n-1}\left(\frac{\partial A}{\partial x_i^{k}}\frac{\partial B}{\partial p_i^{k}}-
\frac{\partial A}{\partial p_i^{k}}\frac{\partial B}{\partial x_i^{k}} +i(-1)^{k}\left(\frac{\overleftarrow{\partial}A}{\partial\psi_i^{k}}\frac{\overrightarrow{\partial}B}{\partial\bar{\psi}_i^{k}}+ \frac{\overleftarrow{\partial}A}{\partial\bar{\psi}_i^{k}}\frac{\overrightarrow{\partial}B}{\partial\psi_i^{k}}\right)\right),
\eea
where $p_i^{k}=(-1)^{k+1}\dot{x}_i^{k}$, and the Hamiltonian
\bea\label{H}
\begin{aligned}
H=\frac{1}{2}\sum_{k=-n+1}^{n-1}(-1)^{k+1}\left(p_i^{k}p_i^{k}+\omega_k^{2}x_i^{k}x_i^{k}+2\omega_k \psi_i^{k}\bar{\psi}_i^{k}\right).
\end{aligned}
\eea
According to (\ref{Poisson}), the canonical structure relations read
\bea\label{canon}
[x_i^{k},p_j^{m}\}=\d_{km}\d_{ij},\qquad [\psi_i^{k},\bar{\psi}_j^{m}\}=i(-1)^{k}\d_{km}\d_{ij}.
\eea

The system (\ref{act2}) holds invariant under the supersymmetry transformations\footnote{Other symmetry transformations of (\ref{act2}) are discussed in Appendix B.}
\bea\label{super}
\d x_i^{k}=\psi_i^{k}\alpha+\bar{\psi}_i^{k}\bar{\alpha}, \qquad \d\psi_i^{k}=-(i\dot{x}_i^{k}-\omega_k x_i^{k})\bar{\alpha}, \qquad \d\bar{\psi}_i^{k}=-(i\dot{x}_i^{k}+\omega_k x_i^{k})\alpha,
\eea
which via the Noether theorem lead to the integrals of motion
\bea\label{Q}
Q=\sum_{k=-n+1}^{n-1}\psi_i^{k}(p_i^{k}-i(-1)^{k+1}\omega_k x_i^{k}),\qquad \bar{Q}=\sum_{k=-n+1}^{n-1}\bar{\psi}_i^{k}(p_i^{k}+i(-1)^{k+1}\omega_k x_i^{k}).
\eea
Along with the Hamiltonian (\ref{H}) the latter obey the structure relations
\bea
[Q,Q\}=0,\quad[H,Q\}=0,\quad [Q,\bar{Q}\}=-2iH,\quad [H,\bar{Q}\}=0,\quad [\bar{Q},\bar{Q}\}=0.
\nonumber
\eea

Let us quantize the model\footnote{Quantization procedure for bosonic infinite-order PU oscillator has been recently considered in paper \cite{Kosinski_1}.} (\ref{act2}). To this end, consider the hermitian bosonic operators $\hat{x}_i^{k}=\left(\hat{x}_i^{k}\right)^{\dag}$, $\hat{p}_i^{k}=\left(p_i^{k}\right)^{\dag}$ and the fermionic operators $\hat{\psi}_i^{k}$, $\hat{\bar{\psi}}_i^{k}$ which are hermitian conjugates of each other $\hat{\bar{\psi}}_i^{k}=\left(\hat{\psi}_i^{k}\right)^\dag$.
In agreement with (\ref{canon}), they
obey the (anti)commutation relations
\bea\label{acr}
[\hat{x}_i^{k},\hat{p}_j^{m}]=i\hbar\d_{km}\d_{ij},\qquad \{\hat{\psi}_i^{k},\hat{\bar{\psi}}_j^{m}\}=(-1)^{k+1}\hbar\d_{km}\d_{ij},
\eea
where $[\cdot,\cdot]$ and $\{\cdot,\cdot\}$ stand for the commutator and anticommutator, respectively.
The quantum Hamiltonian\footnote{We choose the Weyl ordering for the fermions  $\psi_i^{k}\bar{\psi}_i^{k}\;\rightarrow\;\frac{1}{2}\left(\hat{\psi}_i^{k}\hat{\bar{\psi}}_i^{k}- \hat{\bar{\psi}}_i^{k}\hat{\psi}_i^{k}\right)$.} $\hat{H}=(\hat{H})^\dag$ and the supersymmetry generators $\hat{Q}$, $\hat{\bar{Q}}=(\hat{Q})^\dag$ do obey the structure relations (\ref{str}).

Introducing the creation $\bar{a}_i^{k}$, $\bar{c}_i^{k}$ and annihilation $a_i^{k}=(\bar{a}_i^{k})^\dag$, $c_i^{k}=(\bar{c}_i^{k})^\dag$ operators\footnote{Here and in what follows we assume $\omega_0\neq 0$.}
\bea\label{create}
\begin{aligned}
&
a_i^{k}=\sqrt{\frac{|\omega_{k}|}{2\hbar}}\hat{x}_i^{k}+i\frac{1}{\sqrt{2|\omega_{k}|\hbar}}\hat{p}_i^{k},\qquad c_i^{k}=\frac{1}{\sqrt{\hbar}}\hat{\psi}_i^{k},
\\[3pt]
&
\bar{a}_i^{k}=\sqrt{\frac{|\omega_{k}|}{2\hbar}}\hat{x}_i^{k}-i\frac{1}{\sqrt{2|\omega_{k}|\hbar}}\hat{p}_i^{k},\qquad \bar{c}_i^{k}=\frac{1}{\sqrt{\hbar}}\hat{\bar{\psi}}_i^{k},
\end{aligned}
\eea
which obey
\bea\label{canon2}
[a_i^{k},\bar{a}_j^{m}]=\d_{km}\d_{ij},\qquad \{c_i^{k},\bar{c}_j^{m}\}=(-1)^{k+1}\d_{km}\d_{ij}
\eea
one can immediately construct Fock spaces associated with each pair
\begin{align}
&
a_i^{k}|0,k,i\rangle=0, &&    |n,k,i\rangle=\frac{(\bar{a}_i^{k})^n}{\sqrt{n!}}|0,k,i\rangle && \mbox{(no sum)}
\nonumber\\
&
c_i^{k}\mathbf{|0,k,i\rangle}=0, && \mathbf{|n,k,i\rangle}=\bar{c}_i^{k}\mathbf{|0,k,i\rangle}, && \mbox{(no sum)}
\nonumber
\end{align}
the full state space being their tensor product. In Fock space associated with each pair one can introduce the conventional number operators $N_{i}^{k}=\bar{a}_i^{k}a_i^{k}$ (no sum) and $\mathbf{N}_{i}^{k}=\bar{c}_i^{k}c_i^{k}$ (no sum)
\bea
N_{i}^{k}|n,k,i\rangle=n_i^{k}|n,k,i\rangle,\qquad \mathbf{N}_{i}^{k}\mathbf{|n,k,i\rangle}=\mathbf{n_i^{k}|n,k,i\rangle},\qquad\mbox{(no sum)}
\nonumber
\eea
where $n_i^{k}=0,1,2,...$, and $\mathbf{n}_{i}^{k}=0,(-1)^{k+1}$. The latter relation stems from the fact that, if one considers the square of number operator $\mathbf{N}_{i}^{k}=\bar{c}_i^{k}c_i^{k}$ (no sum) and takes into account (\ref{canon2}), one gets
\bea
\mathbf{N}_{i}^{k}\mathbf{N}_{i}^{k}=\bar{c}_i^{k}c_i^{k}\bar{c}_i^{k}c_i^{k}=\bar{c}_i^{k}((-1)^{k+1}-\bar{c}_i^{k}c_i^{k})c_i^{k}=(-1)^{k+1}\mathbf{N}_{i}^{k},
\qquad\mbox{(no sum)}
\nonumber
\eea
from which it follows $\mathbf{N}_{i}^{k}\mathbf{N}_{i}^{k}+(-1)^{k}\mathbf{N}_{i}^{k}=0$ (no sum). This implies that $0$ and $(-1)^{k+1}$ are the eigenvalues of the operator $\mathbf{N}_{i}^{k}$. Moreover, if we assume that $\mathbf{|0,k,i\rangle}$ with even $k$ have positive norms, then $\mathbf{|-1,k,i\rangle}$ are necessarily the negative norm states. Indeed, taking into account (\ref{canon2}) one finds
\bea
\mathbf{\langle-1,k,i|-1,k,i\rangle}=\mathbf{\langle0,k,i|}\,c_i^{k}\,\bar{c}_i^{k}\,\mathbf{|0,k,i\rangle}=-\mathbf{\langle0,k,i|0,k,i\rangle},\qquad\mbox{(no sum).}
\nonumber
\eea
This means that the full state space, which is a tensor product of Fock spaces associated with the ladders, contains negative norm states. We thus conclude that supersymmetry alone can not help to resolve the ghost problem  intrinsic to our system.

Concluding this section, let us analyze the energy spectrum. Rewriting the Hamiltonian (\ref{H}) in terms of the creation and annihilation operators (\ref{create})
\bea
\hat{H}=-\hbar\omega_{0}(\bar{a}_i^{0}a_i^{0}-\bar{c}_i^{0}c_i^{0})+\sum_{k=1}^{n-1}(-1)^{k+1}\hbar\omega_{k}\left(\bar{a}_i^{-k}a_i^{-k}+
\bar{c}_i^{-k}c_i^{-k}+\bar{a}_i^{k}a_i^{k}-\bar{c}_i^{k}c_i^{k}+d\right)
\nonumber
\eea
one can readily compute the energy eignevalues
\bea
E(n_{i}^{k},\mathbf{n}_{i}^{k})=\sum_{i=1}^{d}\left(\sum_{k=1}^{n-1}(-1)^{k+1}\hbar\omega_k(n_{i}^{-k}+\mathbf{n}_{i}^{-k}+n_{i}^{k}-\mathbf{n}_{i}^{k}+d)
-\hbar\omega_0(n_{i}^{0}-\mathbf{n}_{i}^{0})\right).
\nonumber
\eea
Note that this is unbounded from below. If one considers a particular case for which $\omega_0=0$, then one gets a mode corresponding to a free $\,\mathcal{N}=2$ superparticle. The spectrum of this system is continuous but
the negative norm states are still present in the full state space as a consequence of (\ref{acr}). The energy spectrum is unbounded from below as in the former case.

\vskip 0.5cm
\noindent
{\bf 4. Conclusion}
\vskip 0.5cm

To summarize, in this work we have constructed an $\,\mathcal{N}=2$ supersymmetric extension of the PU oscillator for the case of arbitrary frequencies of oscillation. The model was shown to be dynamically equivalent to a set
of decoupled $\,\mathcal{N}=2$ supersymmetric harmonic oscillators with alternating sign in the Hamiltonian. Quantization of the system revealed the presence of
negative norm states and the unbounded from below energy spectrum. The presence of the negative norm states was linked to the fact that the fermionic oscillators enter the action functional with alternating sign.

In a recent work \cite{Reyes}, a fermionic analogue of the PU oscillator has been investigated. It was shown that the corresponding Hamiltonian can be represented as the sum of two fermionic oscillators with alternating sign.
A link to a fermionic sector of the so-called Myers-Pospelov model \cite{Myers} has been established. It would be interesting to study possible relations of the latter model with the system introduced in this work.

Another possible development is to consider the degenerate case in which frequencies of oscillation are equal to each other. As was shown in \cite{Pais}, \cite{Mannheim}-\cite{Smilga_3}, this case requires a more sophisticated construction.

\vskip 0.5cm
\noindent
{\bf Acknowledgements}
\vskip 0.5cm

\noindent
We thank A. Galajinsky for the comments on the manuscript. This work was supported
by the Dynasty Foundation, the MSE program "Nauka" under
the project 3.825.2014/K, RFBR grant 14-02-31139-mol, and the TPU grant LRU.FTI.123.2014.

\vskip 0.8cm
\noindent
{\bf Appendix A. Computation of coefficients}
\vskip 0.5cm
Let us demonstrate how one can obtain the expressions (\ref{coef}). We discuss $\beta_{k}$ in detail. Other coefficients are constructed likewise.

In order to solve the equation
$$
\sum_{m=-n+1}^{n-1}\beta_m\psi_i^{m}=-\psi_i,\eqno{(A1)}
$$
one rewrites the variables $\psi_i^{m}$ in the following form
$$
\psi_i^{m}=\prod_{k=1-n\atop k\neq m}^{n-1}\left(\frac{d}{dt}-i\omega_k\right)\psi_i=
\sum_{k=0}^{2n-2}\sigma_{k,m}\frac{d^k}{dt^k}\psi_i=0,
$$
where we denoted
$$
\sigma_{2n-2,m}\equiv 1,\qquad\sigma_{k,m}= (-i)^{2n-k-2}\sum_{i_1<i_2<..<i_{2n-k-2}\atop i_1,i_2,..,i_{2n-k-2}\neq m}\omega_{i_1}\omega_{i_2}..\omega_{i_{2n-k-2}}.
$$
Then ($A1$) takes a form
$$
\sum_{m=-n+1}^{n-1}\beta_m\sum_{k=0}^{2n-2}\sigma_{k,m}\frac{d^k}{dt^k}\psi_i=-\psi_i.
$$
The coefficient $\beta_m$ is found as a solution of the following matrix equation:
\bea
\left(
\begin{aligned}
&\sigma_{0,-n+1}&& \sigma_{0,-n+2}&& \sigma_{0,-n+3}&&...&& \sigma_{0,n-2}&& \sigma_{0,n-1}\\[2pt]
&\sigma_{1,-n+1}&& \sigma_{1,-n+2}&& \sigma_{1,-n+3}&&...&& \sigma_{1,n-2}&& \sigma_{1,n-1}\\[2pt]
&\sigma_{2,-n+1}&& \sigma_{2,-n+2}&& \sigma_{2,-n+3}&&...&& \sigma_{2,n-2}&& \sigma_{2,n-1}\\[2pt]
&\;\;\;\;....&& .... &&\;\;\;\; .... && .... &&\;\;\;\; .... &&\;\;\;\; ....\\[2pt]
&\sigma_{2n-3,-n+1}&& \sigma_{2n-3,-n+2}&& \sigma_{2n-3,-n+3}&&...&& \sigma_{2n-3,n-2}&& \sigma_{2n-3,n-1}\\[2pt]
&1&&1&&1&&...&&1&&1
\end{aligned}
\right)
\left(
\begin{aligned}
&\beta_{-n+1}\\[2pt]
&\beta_{-n+2}\\[2pt]
&\beta_{-n+3}\\[2pt]
&\;\;\;...\\[2pt]
&\beta_{n-2}\\[2pt]
&\beta_{n-1}
\end{aligned}
\right)=
\left(
\begin{aligned}
&-1\\[2pt]
&\;\,0\\[2pt]
&\;\,0\\[2pt]
&\;...\\[2pt]
&\;\,0\\[2pt]
&\;\,0
\end{aligned}
\right)
\nonumber
\eea
$$
\quad\eqno{(A2)}
$$
Note that
an analogue of this system for the bosonic PU oscillator has been considered in \cite{Pais}.

Denoting the determinant of the matrix entering (A2) by $\Delta(\omega_{-n+1},\omega_{-n+1},..,\omega_{n-1})$, one obtains the recurrence relation
$$
\Delta(\omega_{-n+1},\omega_{-n+2},..,\omega_{n-1})=(-i)^{2n-2}(-1)^{2n}\prod_{i=-n+2}^{n-1}(\omega_{-n+1}-\omega_{i})
\Delta(\omega_{-n+2},\omega_{-n+3},..,\omega_{n-1}), \eqno{(A3)}
$$
which yields
$$
\Delta(\omega_{-n+1},\omega_{-n+2},..,\omega_{n-1})=(-i)^{(n-1)(2n-1)}(-1)^{n-1}\prod_{i_{1}<i_{2}}(\omega_{i_1}-\omega_{i_2}).
$$
In obtaining ($A3$), the identity
$$
\sigma_{k,l_2}-\sigma_{k,l_1}=(\omega_{l_1}-\omega_{l_2})(-i)^{2n-k-2}\sum_{i_1<i_2<..<i_{2n-k-3}\atop i_1,i_2,..,i_{2n-k-3}\neq l_1,l_2}\omega_{i_1}\omega_{i_2}..\omega_{i_{2n-k-3}}
$$
proves to be helpful. If all frequencies are different, the determinant is nonzero.  Moreover, since $\omega_{-k}=-\omega_{k}$, all frequencies must be nonzero but for $\omega_0$ which is allowed to take zero value.

At the next step, one obtains the determinant $\Delta_{m}$ which corresponds to the matrix of the system ($A2$) with $m$-th column being replaced by the column on the right hand side of ($A2$)
\bea
\Delta_{m}=(-1)^{m+1}(-i)^{(n-1)(2n-3)}\prod_{i_1<i_2=-n+1\atop i_1,i_2\neq m}^{n-1}(\omega_{i_1}-\omega_{i_2}).
\nonumber
\eea
The standard Cramer's rule then gives $\beta_{m}$ in (\ref{coef}).

Note that the last equation in ($A2$) $\sum\limits_{k=-n+1}^{n-1}\beta_{k}=0$ implies that any set of unequal frequencies $\omega_{k}$ with $k=0,1,..,n-1$ (not necessarily put in the order of increasing $0\leq\omega_0<\omega_{1}<..<\omega_{n-1}$ as was assumed above) leads to the presence of both positive and negative coefficients $\beta_k$. By this reason, one inevitably reveals negative norm states in quantum theory of the $\,\mathcal{N}=2$ supersymmetric PU oscillator for unequal frequencies.

\vskip 0.5cm
\noindent
{\bf Appendix B. Additional symmetry transformations}
\vskip 0.5cm

For unequal frequencies of oscillation the action functional (\ref{act2}),
apart from being invariant under the time translations and supersymmetry transformations (\ref{tr}), holds invariant under the bosonic and fermionic translations
\bea
\d x_i^{k}=\cos{(\omega_k t)}a_i^{k}+\frac{1}{\omega_k}\sin{(\omega_k t)}b_i^{k},\quad \d\psi_{i}^{k}=e^{it\omega_k}\alpha_i^{k},\quad \d\bar{\psi}_i^{k}=e^{-it\omega_k}\bar{\alpha}_i^{k},
\nonumber
\eea
the rotations
\bea
\d x_i^{k}=\omega_{ij}x_j^{k},\qquad \d\psi_i^{k}=\omega_{ij}\psi_j^{k},\qquad \d\bar{\psi}_i^{k}=\omega_{ij}\bar{\psi}_j^{k},
\nonumber
\eea
with $\omega_{ij}=-\omega_{ji}$, as well as under $U(1)$ $R$-symmetry transformations
\bea
\d\psi_i^{k}=i\nu\psi_i^{k},\qquad \d\bar{\psi}_i^{k}=-i\nu\bar{\psi}_i^{k}.
\nonumber
\eea

The Noether theorem yields integrals of motion
\bea
&&
P_i^{k}=(-1)^{k+1}p_i^{k}\cos{(\omega_k t)}+\omega_k x_i^k \sin{(\omega_k t)},\qquad\quad\; \Psi_i^{k}=e^{-i\omega_k t}\psi_i^{k},
\nonumber
\\[6pt]
&&
X_i^{k}=-\frac{1}{\omega_k} p_i^{k} \sin{(\omega_k t)}+(-1)^{k+1}x_i^{k}\cos{(\omega_k t)},\quad\;\;\;\;\;\, \bar{\Psi}_i^{k}=e^{i\omega_k t}\bar{\psi}_i,
\nonumber
\\[2pt]
&&
M_{ij}=\sum_{k=-n+1}^{n-1}\left(-x^{k}_{[i}p^{k}_{j]}+i(-1)^{k+1}\psi^{k}_{[i}\bar{\psi}^{k}_{j]}\right),\qquad J=\sum_{k=-n+1}^{n-1}(-1)^{k+1}\psi_i^{k}\bar{\psi}_i^{k}.
\nonumber
\eea

Under the graded Poisson bracket (\ref{Poisson}) these conserved charges along with the Hamiltonian (\ref{H}) and the supercharges (\ref{Q}) obey the following non-vanishing (anti)commutation relations:
\bea
\begin{aligned}
&
[Q,\bar{Q}\}=-2iH,\qquad\quad[X_i^{k},P_j^{m}\}=\d_{ij}\d_{km},\qquad\quad [\Psi_i^{k},\bar{\Psi}_j^{m}\}=i(-1)^{k}\d_{ij}\d_{km},
\\[5pt]
&
[H,P_i^{k}\}=(-1)^{k+1}\omega_k^{2}X_i^{k},\quad [H,\Psi_i^{k}\}=-i\omega_{k}\Psi_i^{k},\quad [J,Q\}=-i Q,\quad[J,\Psi_i^{k}\}=-i\Psi_i^{k},
\\[5pt]
&
[H,X_i^{k}\}=(-1)^{k}P_i^{k},\qquad\quad [H,\bar{\Psi}_i^{k}\}=i\omega_{k}\bar{\Psi}_i^{k},\quad\;\;\, [J,\bar{Q}\}=i\bar{Q},\quad\;\;\;
 [J,\bar{\Psi}_i^{k}\}=i\bar{\Psi}_i^{k},
\\[5pt]
&
[Q,\bar{\Psi}_i^{k}\}=-i P_i^{k}+(-1)^{k}\omega_k X_i^{k},\quad [Q,P_i^{k}\}=-i\omega_k \Psi_i^{k},\quad [Q,X_i^{k}\}=(-1)^{k}\Psi_i^{k},
\\[5pt]
&
[\bar{Q},\Psi_i^{k}\}=-i P_i^{k}-(-1)^{k}\omega_k X_i^{k},\quad [\bar{Q},P_i^{k}\}=i\omega_{k}\bar{\Psi}_i^{k},\quad\;\;\, [\bar{Q},X_i^{k}\}=(-1)^{k}\bar{\Psi}_i^{k}, \quad
\nonumber
\\[5pt]
&
[M_{ij},A_s^{k}\}=A_i^{k}\d_{js}-A_{j}^{s}\d_{is},
\end{aligned}
\nonumber
\eea
where $A_s^{k}=P_s^{k},\,X_s^{k},\,\Psi_s^{k},\,\bar{\Psi}_s^{k}$.

\end{document}